\documentclass[10pt,aps,pra,superscriptaddress,twocolumn,floatfix]{revtex4-2}

\usepackage[utf8]{inputenc}
\usepackage[english]{babel}
\usepackage[none]{hyphenat}
\usepackage{graphicx}
\usepackage{amsmath}
\usepackage{amssymb}
\usepackage{mathtools}
\usepackage{bbm}
\usepackage{hyperref}
\usepackage{nicematrix}
\usepackage{braket}
\usepackage[normalem]{ulem}
\usepackage{float}
\usepackage[export]{adjustbox}
\usepackage[ruled, vlined, onelanguage]{algorithm2e}
\usepackage{booktabs} 
\usepackage{siunitx} 
\usepackage[justification=raggedright,singlelinecheck=false]{caption}

\hypersetup{colorlinks=true, linkcolor=blue, citecolor=blue, urlcolor=blue}

\definecolor{headercolor}{RGB}{94, 89, 72}    
\definecolor{rowcolor1}{RGB}{245, 244, 242}   
\definecolor{rowcolor2}{RGB}{255, 255, 255}   
\definecolor{bestcolor}{RGB}{229, 229, 233}   
\definecolor{bestvaluecolor}{RGB}{28, 42, 67} 
\definecolor{ourscolor}{RGB}{198, 193, 165}   

\begin{document}
\title{QUIET-SR: Quantum Image Enhancement Transformer for Single Image Super-Resolution}

\author{\textbf{Siddhant Dutta}}
\email{siddhant010@e.ntu.edu.sg}
\affiliation{College of Computing \& Data Science, Nanyang Technological University (NTU),
Singapore, 639798, Singapore}
\affiliation{SVKM's Dwarkadas J. Sanghvi College of Engineering, Mumbai, India}

\author{\textbf{Nouhaila Innan}}
\email{nouhaila.innan@nyu.edu}
\affiliation{
eBRAIN Lab, Division of Engineering, New York University Abu Dhabi (NYUAD), Abu Dhabi, UAE}
\affiliation{Center for Quantum and Topological Systems (CQTS), NYUAD Research Institute, NYUAD, Abu Dhabi, UAE}

\author{\textbf{Khadijeh Najafi}}
\email{sonaa.najafi@gmail.com}
\affiliation{IBM Quantum, IBM T.J. Watson Research Center, Yorktown Heights, 10598, USA}
\affiliation{MIT-IBM Watson AI Lab, Cambridge MA, 02142, USA}

\author{\textbf{Sadok Ben Yahia}}
\email{say@mmmi.sdu.dk}

\affiliation{The Maersk Mc-Kinney Moller Institute, University of Southern Denmark, Sønderborg, Denmark}
\affiliation{Department of Software Science, Tallinn University of Technology, Tallinn, Estonia}

\author{\textbf{Muhammad Shafique}}
\email{muhammad.shafique@nyu.edu}

\affiliation{
eBRAIN Lab, Division of Engineering, New York University Abu Dhabi (NYUAD), Abu Dhabi, UAE}
\affiliation{Center for Quantum and Topological Systems (CQTS), NYUAD Research Institute, NYUAD, Abu Dhabi, UAE}

\begin{abstract}
Recent advancements in Single-Image Super-Resolution (SISR) using deep learning have significantly improved image restoration quality. However, the high computational cost of processing high-resolution images due to the large number of parameters in classical models, along with the scalability challenges of quantum algorithms for image processing, remains a major obstacle. In this paper, we propose the Quantum Image Enhancement Transformer for Super-Resolution (QUIET-SR), a hybrid framework that extends the Swin transformer architecture with a novel shifted quantum window attention mechanism, built upon variational quantum neural networks. QUIET-SR effectively captures complex residual mappings between low-resolution and high-resolution images, leveraging quantum attention mechanisms to enhance feature extraction and image restoration while requiring a minimal number of qubits, making it suitable for the Noisy Intermediate-Scale Quantum (NISQ) era. We evaluate our framework in MNIST (30.24 PSNR, 0.989 SSIM), FashionMNIST (29.76 PSNR, 0.976 SSIM) and the MedMNIST dataset collection, demonstrating that QUIET-SR achieves PSNR and SSIM scores comparable to state-of-the-art methods while using fewer parameters. Our efficient batching strategy directly enables massive parallelization on multiple QPU's paving the way for practical quantum-enhanced image super-resolution through coordinated QPU–GPU quantum supercomputing.
\end{abstract}
\maketitle
\section{Introduction}
Single-Image Super-Resolution (SISR) aims to recover a High-Resolution (HR) image from a Low-Resolution (LR) input image \cite{chen2024single}. Formally, given a low-resolution image \( I_{LR} \), the objective of SISR is to reconstruct a high-resolution image \( I_{HR} \) such that the perceptual quality and structural fidelity of the resulting image closely approximate those of an ideal high-resolution reference. More formally, the original high-resolution image can be considered the ground truth, representing the scene sampled at a sufficiently high spatial resolution to capture fine details without degradation. In a physical sense, real-world scenes exist as continuous signals, and an ideal high-resolution image corresponds to a discretized representation with minimal loss of spatial information. The aim is to learn a mapping \( f: I_{LR} \rightarrow I_{HR} \) that reconstructs an image \( I_{HR} \) approximating this ideal ground truth. This technique has significant applications in fields such as medical imaging, satellite imagery, and autonomous driving as it helps enhance scans like MRIs, aiding in disease detection, improves the clarity of earth observation data for environmental monitoring and disaster response, and sharpens camera inputs, improving object recognition for safer navigation respectively. By recovering lost details, SISR enables better decision-making across these fields \cite{rs14215423,7095900,10.1007/978-3-319-10662-5_18}.

In the context of deep learning, traditional SISR methods often use Convolutional Neural Networks (CNNs) that incorporate residual learning, a technique that helps the network focus on reconstructing lost high-frequency details rather than relearning the entire image structure \cite{dong2015image,lim2017enhanced}. Another common approach involves Generative Adversarial Networks (GANs), which generate high-resolution images by training two competing networks to improve image quality \cite{ledig2017photo}. However, these methods are computationally demanding and struggle to scale efficiently as the size of the image and complexity increase. Recent developments in SISR models, such as those based on Swin Transformer architectures, have demonstrated state-of-the-art performance in capturing fine details while maintaining contextual integrity, highlighting the ongoing demand for scalable and efficient approaches in this field \cite{liang2021swinir, conde2022swin2sr}.
\begin{figure*}[htbp]
    \centering
    \includegraphics[width=1\linewidth]{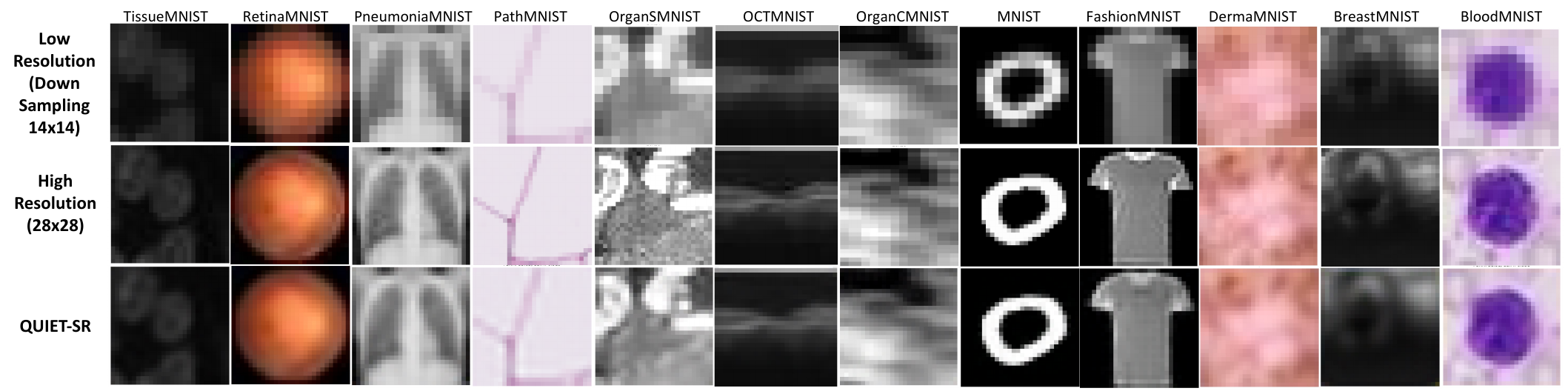}
    \caption{ \small This grid compares low-resolution (14×14), high-resolution (28×28), and QUIET-SR super-resolution images across multiple MNIST-like datasets. The high-resolution row represents the ground truth, while the QUIET-SR row demonstrates the model's capability to reconstruct fine details and preserve structural integrity. The super-resolved images generated by QUIET-SR closely approximate the high-resolution ground truth, effectively enhancing image clarity and preserving key features.}
    \label{fig:comparision}
\end{figure*}
Quantum Computing (QC) introduces a representational-based computational paradigm that leverages quantum states to encode and manipulate high-dimensional data more efficiently. This approach has the potential to overcome certain limitations of classical computation in tasks requiring complex feature representations, such as extracting fine textures, spatial relationships, and high-frequency details in image processing \cite{wang2022review}. Classical methods often struggle with the exponential growth of data, high memory and computational demands, and the difficulty of solving non-convex optimization problems, making quantum techniques a promising alternative for more scalable and efficient solutions. The field of Quantum Machine Learning (QML) utilizes the unique properties of quantum mechanics to investigate state spaces that are infeasible via classical approaches \cite{innan2024financial, biamonte2017quantum}, to improve learning efficiency and model generalization. However, practical implementations of quantum image processing face challenges, primarily due to limitations in qubit resources and the presence of quantum noise \cite{wang2022review,senokosov2024quantum}. Current research in quantum Super-Resolution (SR) predominantly focuses on theoretical frameworks with limited practical implementations. This is largely due to the qubit requirements that exceed the capabilities of existing hardware \cite{unlu2024hybrid, comajoan2024quantum}. Several works have employed adiabatic QC with D-Wave systems, achieving proof-of-concept results \cite{choong2023quantum}.

In this paper, we tackle the dual challenge of leveraging Noisy Intermediate-Scale Quantum (NISQ) computers \cite{preskill2018quantum} while efficiently integrating QC with classical deep learning for super-resolution. A major obstacle is the inherent difficulty of combining quantum feature representations with classical deep learning models in a way that effectively utilizes the potential quantum advantages without being bottlenecked by hardware limitations. This challenge motivates our exploration of NISQ devices and the development of hybrid approaches that can overcome their constraints while achieving comparable or improved performance in SISR. To address this, we introduce a novel hybrid quantum-classical architecture for SISR. Our proof-of-concept framework demonstrates a scalable hybrid quantum super-resolution system, operating with fewer than 10 qubits per circuit within the proposed architecture. This represents a step toward practical quantum applications in image processing while staying within current quantum hardware constraints, as shown in Fig. \ref{fig:comparision}.

\begin{figure*}[t]
    \centering
    \includegraphics[width=1\linewidth]{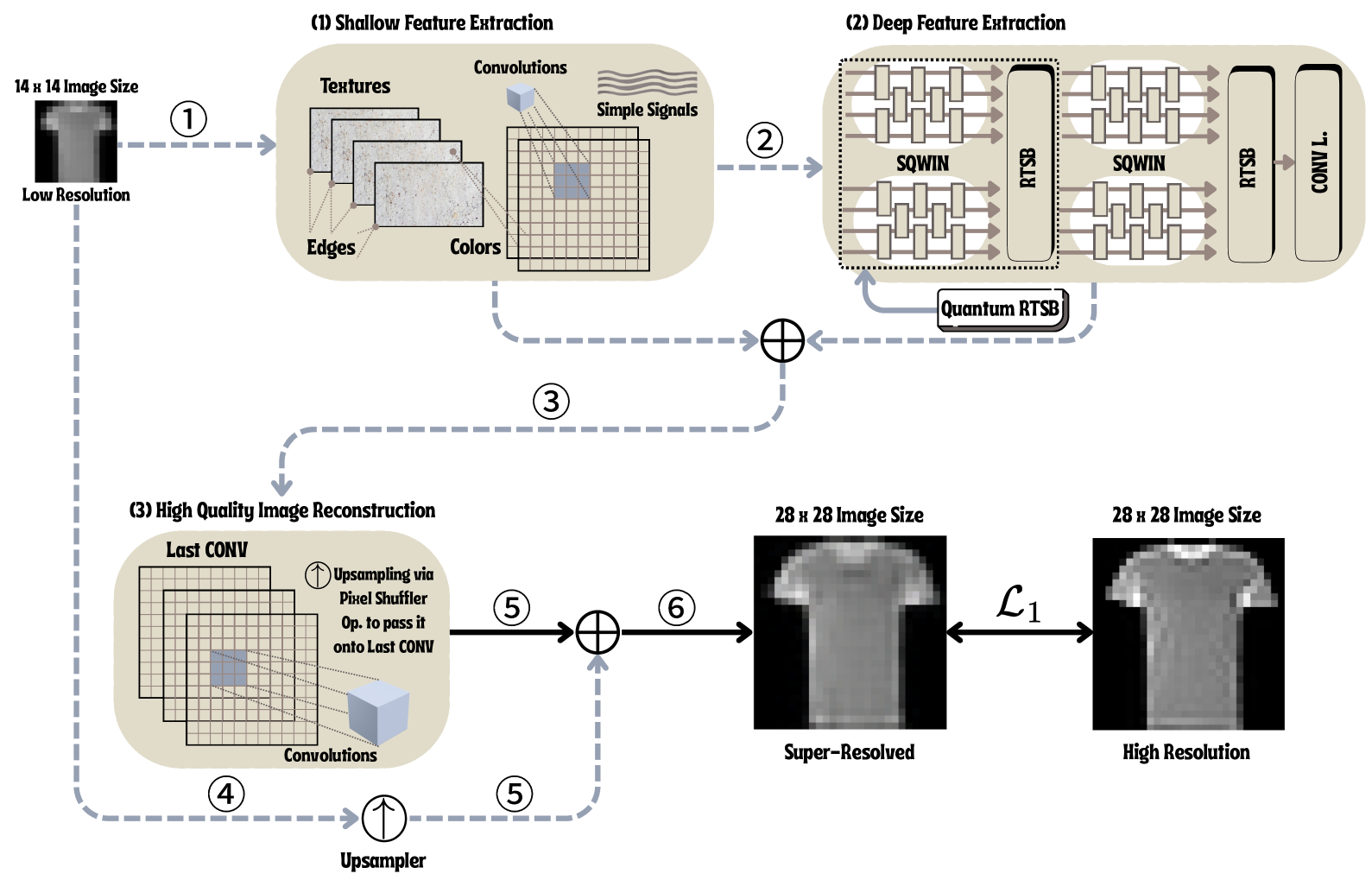}
    \caption{ \small \textbf{High-level workflow of the QUIET-SR framework.} The architecture processes a $14 \times 14$ Low-Resolution (LR) input through three main stages: \textbf{(1) Shallow Feature Extraction}, where initial convolutional layers capture low-frequency information such as edges, textures, and colors; \textbf{(2) Deep Feature Extraction}, which utilizes Quantum Residual Transformer Blocks (Quantum RTSB) containing the specific \textbf{SQWIN} (Shifted Quantum Window) attention mechanism to model complex dependencies; and \textbf{(3) High Quality Image Reconstruction}, where features are aggregated via global residual connections surmised via \textbf{(4-5)} upsampling via a Pixel Shuffle operation and final convolutions to synthesize the $28 \times 28$ Super-Resolved output. \textbf{(6)} The network is optimized by minimizing the $\mathcal{L}_1$ loss between the generated image and the High-Resolution ground truth. The QNN architecture uses angle embedding and basic entangler layers to transform features and optimize attention efficiency. The final stage, SR Image Reconstruction, synthesizes the SR output, demonstrating the advantages of quantum-enhanced image restoration}
    \label{overview}
\end{figure*}

Our work as shown \ref{overview} presents the 1$^{st}$ resource-aware quantum attention module for image SR, preserving Swin-style shifted-window locality \& running on today's $\leq$10-qubit devices. Prior quantum vision approaches either (i) use non-scalable encodings, (ii) address only classification, or (iii) rely on quantum annealing. In contrast, our gate-based, NISQ-compatible design directly tackles SR, marking the  1$^{st}$ quantum realization of shifted-window attention.

\noindent \textbf{Significance for the CV community:}
SQWIN, as a technique is the  1$^{st}$ scalable NISQ compatible method that bridges QC \& CV. It uses quantum-native shifted window attention to enable entangled cross-window interactions \& complex-valued scores, going beyond classical layers. 

\noindent \textbf{Significance for the QC community:}
This is the  1$^{st}$ gate-based quantum analysis of a structured SR task, quantifying the expressivity of quantum attention modules \& offering hardware-relevant benchmarks for future quantum design.

\textbf{The novel contributions of our work can be summarized as follows:}

\begin{enumerate}
    \item We introduce QUIET-SR, a novel hybrid quantum-classical model that enhances image resolution. Our approach integrates a quantum-based attention mechanism with a classical deep learning framework for image reconstruction. The quantum component helps efficiently capture important image features, while the classical deep network ensures high-quality output.
    \item We demonstrate that Quantum Neural Networks (QNNs) can be effectively used for image super-resolution within current hardware constraints. Our model operates using fewer than 10 qubits per circuit, making it the first functional variational quantum approach that is feasible on today's quantum devices.
    \item We propose Shifted Quantum Window Attention (SQWIN), a novel quantum attention mechanism that processes images by dividing them into small, non-overlapping regions (fixed-size windows) and shifting them in a structured way to capture spatial relationships more effectively. Using quantum states to process these windows, SQWIN improves the ability of the model to capture fine image details while maintaining scalability on near-term quantum hardware and paving the way toward practical image super-resolution enabled by coordinated QPU-GPU quantum supercomputing.
\end{enumerate}
The rest of the paper is organized as follows, Sec. \ref{sec2} reviews related work in deep learning-based and quantum-enhanced super-resolution; Sec. \ref{sec3} presents the proposed QUIET-SR framework, detailing patch embedding, SQWIN, and complexity analysis; Sec. \ref{sec4} describes the experimental setup, benchmarking models, and performance evaluation using PSNR and SSIM; Sec. \ref{sec5} concludes with key findings, limitations, and future directions for hybrid quantum-classical approaches in image super-resolution.
\section{Experiments and Results \label{sec4}}
\begin{table*}[t]
    \centering
    \small
    \setlength{\tabcolsep}{4pt}
    \begin{minipage}{\textwidth}
        \caption{Quantitative comparisons of PSNR/SSIM across six complicated medical image datasets for an embedding dimension equal to the number of qubits (4). The best values among our models are highlighted in bold green, while the closest best values from current standard SOTA models are highlighted in a lighter green (Part 1).}
        \resizebox{\textwidth}{!}{%
        \begin{NiceTabular}{l|cc|cc|cc|cc|cc|cc}[colortbl-like]
            \toprule
            \rowcolor{headercolor} 
            \textcolor{white}{\textbf{Method}} & 
            \multicolumn{2}{c|}{\textcolor{white}{\textbf{BloodMNIST}}} & 
            \multicolumn{2}{c|}{\textcolor{white}{\textbf{BreastMNIST}}} & 
            \multicolumn{2}{c|}{\textcolor{white}{\textbf{DermaMNIST}}} & 
            \multicolumn{2}{c|}{\textcolor{white}{\textbf{OCTMNIST}}} & 
            \multicolumn{2}{c|}{\textcolor{white}{\textbf{OrganCMNIST}}} & 
            \multicolumn{2}{c}{\textcolor{white}{\textbf{OrganSMNIST}}} \\
            \rowcolor{headercolor} & 
            \textcolor{white}{PSNR} & \textcolor{white}{SSIM} & 
            \textcolor{white}{PSNR} & \textcolor{white}{SSIM} & 
            \textcolor{white}{PSNR} & \textcolor{white}{SSIM} & 
            \textcolor{white}{PSNR} & \textcolor{white}{SSIM} & 
            \textcolor{white}{PSNR} & \textcolor{white}{SSIM} & 
            \textcolor{white}{PSNR} &  \textcolor{white}{SSIM} \\
            \midrule
            \rowcolor{rowcolor1} Nearest Neighbor & 18.35 & 0.748 & 16.46 & 0.689 & 20.35 & 0.812 & 19.35 & 0.792 & 12.95 & 0.594 & 12.93 & 0.611 \\
            \rowcolor{rowcolor2} Bilinear & 19.87 & 0.772 & 17.98 & 0.712 & 21.97 & 0.834 & 20.83 & 0.816 & 14.38 & 0.625 & 14.35 & 0.642 \\
            \rowcolor{rowcolor1} Bicubic & 21.63 & 0.812 & 19.71 & 0.752 & 22.72 & 0.867 & 22.58 & 0.848 & 16.14 & 0.662 & 16.08 & 0.683 \\
            \rowcolor{rowcolor2} Sparse Representation & 22.15 & 0.836 & 20.23 & 0.776 & 22.84 & 0.891 & 22.81 & 0.874 & 17.69 & 0.683 & 17.63 & 0.705 \\
            \rowcolor{rowcolor1} Iterative Back-projection & 22.68 & 0.861 & 21.75 & 0.799 & 22.96 & 0.913 & 22.94 & 0.896 & 19.24 & 0.709 & 19.18 & 0.731 \\
            \rowcolor{rowcolor2} SRCNN & 29.20 & 0.896 & 26.27 & 0.832 & 36.28 & 0.941 & 31.17 & 0.928 & 20.79 & 0.742 & 20.73 & 0.764 \\
            \rowcolor{bestcolor} Swin2SR & 30.42 & 0.932 & 27.49 & 0.872 & 37.55 & 0.961 & 32.44 & 0.949 & 21.93 & 0.768 & 21.89 & 0.792 \\
            \rowcolor{ourscolor} \textbf{QUIET-SR} (ours) & \textcolor{bestvaluecolor}{\textbf{31.24}} & \textcolor{bestvaluecolor}{\textbf{0.950}} & \textcolor{bestvaluecolor}{\textbf{28.35}} & \textcolor{bestvaluecolor}{\textbf{0.894}} & \textcolor{bestvaluecolor}{\textbf{38.24}} & \textcolor{bestvaluecolor}{\textbf{0.973}} & \textcolor{bestvaluecolor}{\textbf{33.24}} & \textcolor{bestvaluecolor}{\textbf{0.963}} & \textcolor{bestvaluecolor}{\textbf{22.80}} & \textcolor{bestvaluecolor}{\textbf{0.814}} & \textcolor{bestvaluecolor}{\textbf{22.81}} & \textcolor{bestvaluecolor}{\textbf{0.811}} \\
            \bottomrule
        \end{NiceTabular}
        }
        \label{tab:medical-comparison1}
        \vspace{1em}
        \caption{Quantitative comparisons of PSNR/SSIM across six datasets from digits, fashion, and medical categories for an embedding dimension equal to the number of qubits (4). The best values among our models are highlighted in bold green, while the closest best values from current SOTA models are highlighted in a lighter green (Part 2).}
        \resizebox{\textwidth}{!}{%
        \begin{NiceTabular}{l|cc|cc|cc|cc|cc|cc}[colortbl-like]
            \toprule
            \rowcolor{headercolor} 
            \textcolor{white}{\textbf{Method}} & 
            \multicolumn{2}{c|}{\textcolor{white}{\textbf{PathMNIST}}} & 
            \multicolumn{2}{c|}{\textcolor{white}{\textbf{PneumoniaMNIST}}} & 
            \multicolumn{2}{c|}{\textcolor{white}{\textbf{RetinaMNIST}}} & 
            \multicolumn{2}{c|}{\textcolor{white}{\textbf{FashionMNIST}}} & 
            \multicolumn{2}{c|}{\textcolor{white}{\textbf{MNIST}}} & 
            \multicolumn{2}{c}{\textcolor{white}{\textbf{TissueMNIST}}} \\
            \rowcolor{headercolor} & 
            \textcolor{white}{PSNR} & \textcolor{white}{SSIM} & 
            \textcolor{white}{PSNR} & \textcolor{white}{SSIM} & 
            \textcolor{white}{PSNR} & \textcolor{white}{SSIM} & 
            \textcolor{white}{PSNR} & \textcolor{white}{SSIM} & 
            \textcolor{white}{PSNR} & \textcolor{white}{SSIM} & 
            \textcolor{white}{PSNR} &  \textcolor{white}{SSIM} \\
            \midrule
            \rowcolor{rowcolor1} Nearest Neighbor & 16.87 & 0.620 & 18.82 & 0.775 & 19.98 & 0.777 & 16.83 & 0.776 & 17.32 & 0.789 & 21.18 & 0.775 \\
            \rowcolor{rowcolor2} Bilinear & 18.39 & 0.651 & 20.34 & 0.806 & 21.50 & 0.808 & 18.35 & 0.807 & 18.84 & 0.820 & 22.70 & 0.806 \\
            \rowcolor{rowcolor1} Bicubic & 20.15 & 0.692 & 21.11 & 0.846 & 22.27 & 0.848 & 20.11 & 0.847 & 20.60 & 0.859 & 22.86 & 0.846 \\
            \rowcolor{rowcolor2} Sparse Representation & 21.67 & 0.721 & 22.63 & 0.876 & 22.79 & 0.878 & 21.63 & 0.877 & 22.12 & 0.889 & 22.93 & 0.876 \\
            \rowcolor{rowcolor1} Iterative Back-projection & 22.19 & 0.751 & 22.75 & 0.906 & 22.86 & 0.908 & 22.15 & 0.907 & 22.64 & 0.919 & 22.98 & 0.906 \\
            \rowcolor{rowcolor2} SRCNN & 26.71 & 0.781 & 30.67 & 0.935 & 31.83 & 0.937 & 27.67 & 0.936 & 28.16 & 0.949 & 35.02 & 0.935 \\
            \rowcolor{bestcolor} Swin2SR & 27.93 & 0.805 & 31.89 & 0.954 & 33.05 & 0.956 & 28.89 & 0.960 & 29.38 & 0.972 & 36.29 & 0.954 \\
            \rowcolor{ourscolor} \textbf{QUIET-SR} (ours) & \textcolor{bestvaluecolor}{\textbf{28.82}} & \textcolor{bestvaluecolor}{\textbf{0.820}} & \textcolor{bestvaluecolor}{\textbf{32.73}} & \textcolor{bestvaluecolor}{\textbf{0.966}} & \textcolor{bestvaluecolor}{\textbf{33.91}} & \textcolor{bestvaluecolor}{\textbf{0.967}} & \textcolor{bestvaluecolor}{\textbf{29.76}} & \textcolor{bestvaluecolor}{\textbf{0.976}} & \textcolor{bestvaluecolor}{\textbf{30.24}} & \textcolor{bestvaluecolor}{\textbf{0.989}} & \textcolor{bestvaluecolor}{\textbf{37.12}} & \textcolor{bestvaluecolor}{\textbf{0.966}} \\
            \bottomrule
        \end{NiceTabular}
        }
        \label{tab:medical-comparison2}
    \end{minipage}
\end{table*}
In this section, we go over the results of QUIET-SR and provide an analysis of the key elements of the design.
\textbf{Datasets.} QUIET-SR is evaluated on MNIST and 12 additional datasets from the MNIST-like family, covering both general and medical imaging domains. To introduce additional complexity, the model is trained on the full dataset without prior knowledge of class information, ensuring generalization without explicit supervision. High-resolution images are downsampled by a factor of 2 to produce low-resolution counterparts with dimensions of 14×14, following standard super-resolution benchmarks. Our evaluation utilizes diverse datasets including MNIST (handwritten digits) \cite{deng2012mnist}, FashionMNIST (clothing/accessories) \cite{xiao2017fashion}, and medical imaging collections \cite{medmnistv2} spanning BloodMNIST (healthy blood cells), BreastMNIST (ultrasounds), DermaMNIST (skin lesions), OCTMNIST (retinal images), OrganSMNIST/CMNIST (liver tumor CT scans), PathMNIST (colorectal cancer histology), PneumoniaMNIST (chest X-rays), RetinaMNIST (fundus images), and TissueMNIST (human tissue histology) to thoroughly assess performance across standard and specialized domains. 

\textbf{Training Setup. }The QUIET-SR architecture consists of six layers, each with a window size of 2, an embedding dimension of 4, and 4 attention heads. A $QMLP$ with a ratio of 2 is used to refine feature representations. The pixel shuffle technique, specifically the auxiliary variant, is employed for upsampling, ensuring efficient reconstruction of fine details. The model is trained using the Adam optimizer \cite{kingma2014adam} with a learning rate of 2$\times$10$^{-4}$, and the $L_1$ loss function \cite{he2022revisiting}, also known as Mean Absolute Error (MAE), is used to minimize the pixel-wise reconstruction error between the super-resolved image $I_{SR}$ and the ground truth high-resolution image $I_{HR}$. It is defined as:
\begin{equation}
    L_1 = \frac{1}{N} \sum_{i=1}^{N} \left| I_{SR}[i] - I_{HR}[i] \right|,
\end{equation}
where $N$ represents the total number of pixels in the image. This loss function penalizes the absolute differences between the predicted and ground truth pixel values, encouraging sharp and accurate reconstructions. Training is conducted for 25 epochs with a batch size of 64, utilizing Ampere GPU acceleration to expedite computation.

\begin{figure*}[htbp]
    \centering
    \begin{minipage}{0.48\linewidth}
        \centering
        \includegraphics[width=\linewidth]{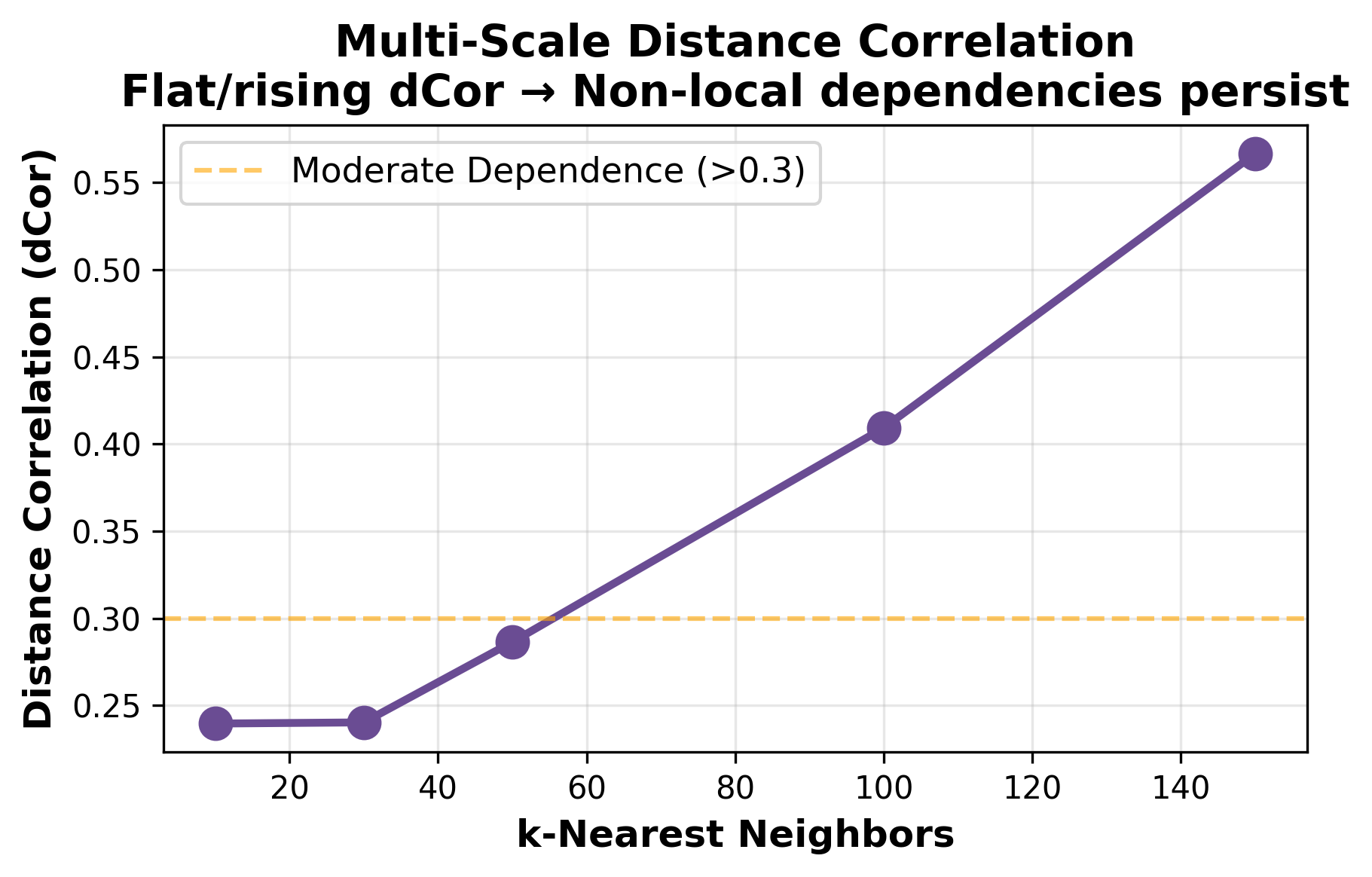}
        \caption*{\textbf{(a) Distance Correlation:} dCor rises with neighborhood size, showing strong long-range correlations.}
    \end{minipage}
    \hfill
    \begin{minipage}{0.48\linewidth}
        \centering
        \includegraphics[width=\linewidth]{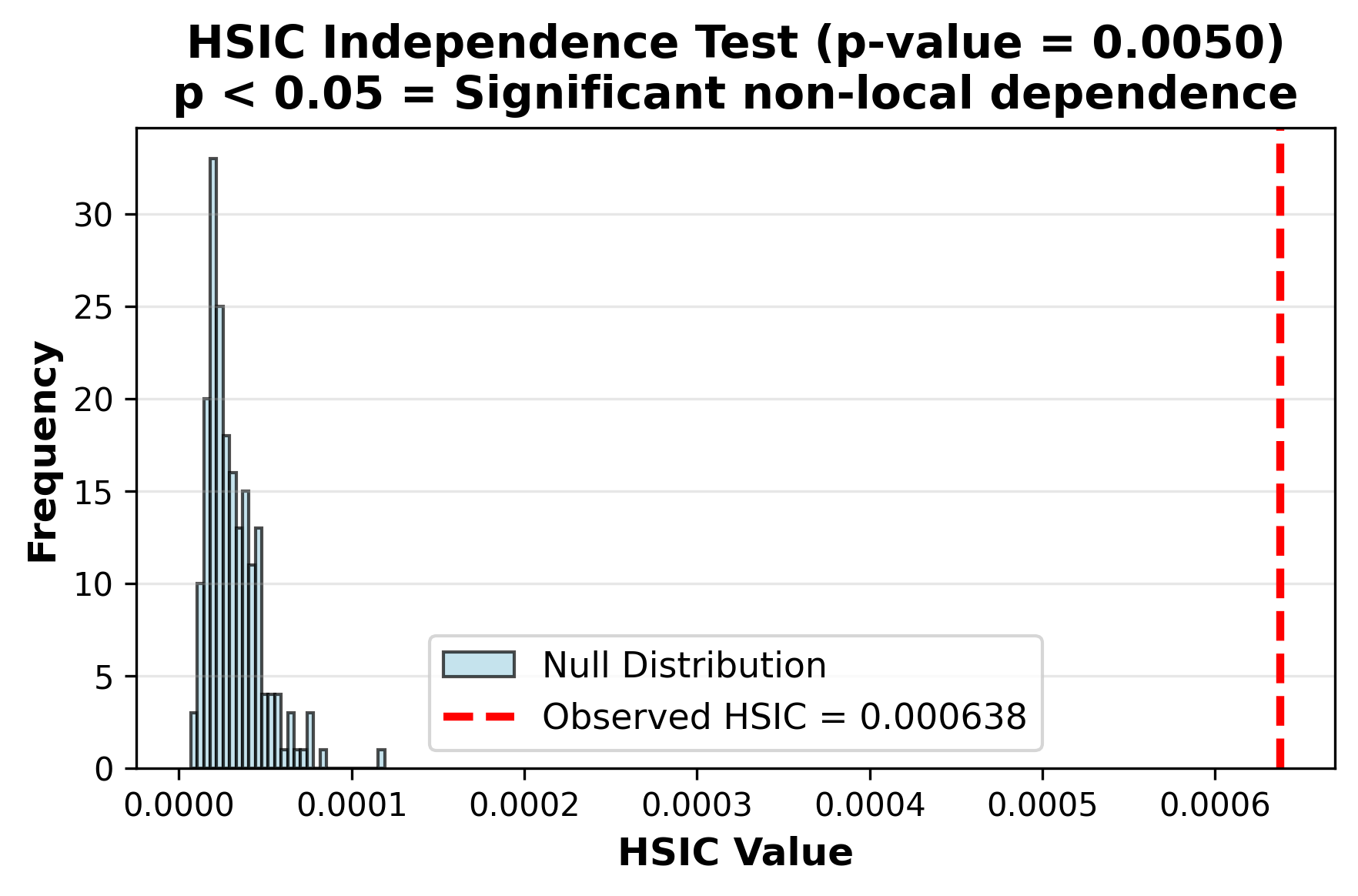}
        \caption*{\textbf{(b) HSIC Independence Test:} Observed statistic (red dashed line) lies far in the tail of the null distribution, confirming significant non-local dependence.}
    \end{minipage}
    \caption{\textbf{Non-local dependence in learned feature space.} (a) Multi-scale Distance Correlation reveals that distant features remain statistically dependent. (b) HSIC testing confirms that the global structure of the embeddings is preserved. Together, these analyses highlight the model's ability to encode long-range correlations crucial for image reconstruction.}
    \label{fig:dcor_hsic}
\end{figure*}

\textbf{Metrics \& Benchmarking Models. }The quality of the reconstructed images is evaluated using two widely adopted metrics: Peak Signal-to-Noise Ratio (PSNR) \cite{6263880} and Structural Similarity Index Measure (SSIM) \cite{1284395}. These metrics quantitatively assess the fidelity of the super-resolved images relative to the ground truth. PSNR measures the logarithmic ratio between the maximum possible signal value and the distortion introduced by reconstruction errors. SSIM quantifies the structural similarity between the super-resolved and ground truth images by considering luminance, contrast, and structure. Higher PSNR \& SSIM values correspond to better reconstruction quality. The benchmarking results can be seen in Tables \ref{tab:medical-comparison1} and \ref{tab:medical-comparison2}.

\subsection*{Clarifications on Experimental Design}
Our approach builds on Swin's shifted window attention but is not a direct copy of any specific variant; the customized implementation is available in the supplementary material. We focused primarily on Swin2SR, as QUIET-SR's core contribution is methodological—introducing the 1$^{st}$ NISQ-compatible quantum attention module for image SR. Rather than attempting to outperform classical SOTA, we aimed to isolate the quantum attention's impact by comparing it to its classical counterpart. No gate-based quantum methods exist for direct comparison (including Choong et al.'s annealing approach). Our experiments validate SQWIN as the 1$^{st}$ scalable quantum-classical SR approach, laying groundwork for future quantum vision research over competing with optimized classical models.

To ensure a rigorous evaluation of the representational efficiency of our proposed quantum attention mechanism, we established a strict resource-constrained baseline for the classical Swin2SR model. Standard Swin2SR implementations utilize large embedding dimensions, resulting in parameter counts that dwarf the capacity of current NISQ-compatible quantum circuits. To isolate the algorithmic advantage of SQWIN from mere scaling benefits, we downscaled the classical Swin2SR embedding dimension to match the qubit count of our quantum layers. This deliberate restriction effectively reduces the classical model to the same informational bottleneck faced by the quantum model. Consequently, any performance parity or advantage observed in QUIET-SR can be directly attributed to the superior expressivity and high dimensional interaction of quantum entanglement in the feature space, rather than discrepancies in model capacity.

In Fig. \ref{framework}, log-CPB refers to the log-spaced continuous relative position bias \cite{conde2022swin2sr}. Fig. \ref{potential} normalizes representational capacity rather than computational cost, since FLOPs are inapplicable to quantum circuits. Here, \textit{Embedding dimension} indicates the latent vector length entering the attention block. Swin2SR is the classical SR baseline.
\vspace{-10pt}
\subsection*{Quantum Feature Representation Analysis}

We evaluated the capacity of the model to capture non-local, long-range correlations in feature space using Distance Correlation (dCor) \cite{szekely2007measuring} and the Hilbert-Schmidt Independence Criterion (HSIC) \cite{gretton2005measuring}. These metrics quantify dependencies beyond local neighborhoods, revealing the global structure of the learned embeddings.

Multi-scale dCor analysis shows a steady increase with neighborhood size ($k$), reaching a pronounced dependence ($\text{dCor} > 0.30$) at $k=150$. This indicates that features far apart in the embedding space remain strongly correlated, highlighting the model's ability to encode long-range interactions critical for high-fidelity image reconstruction.

HSIC testing confirms these findings. The observed statistic of $0.000638$ lies far in the tail of the null distribution, with a p-value well below $0.05$, indicating highly significant non-local dependence. Together, these results demonstrate that the feature manifold preserves both local and global structure.

\subsection*{QUIET-SR Key Results and SOTA Comparison }The quantitative evaluation suggests that QUIET-SR demonstrates improved performance metrics compared to conventional approaches, with enhanced PSNR indicating reduced pixel-wise reconstruction errors and higher SSIM values suggesting better preservation of structural information in the reconstructed images as shown in Fig. \ref{fig:comparision}. As evidenced in the Tables \ref{tab:medical-comparison1} and \ref{tab:medical-comparison2}, conventional interpolation techniques yield comparatively lower scores due to their limited capacity to reconstruct fine details, while bicubic interpolation shows only marginal improvements but exhibits limitations in preserving complex textural information. Methods employing iterative back-projection and sparse representation demonstrate incremental improvements through frequency-domain analysis and iterative refinement processes; however, their performance metrics remain below those achieved by deep learning architectures, with SRCNN achieving improved metrics through hierarchical feature learning despite being constrained by architectural depth limitations, and Swin2SR showing notable improvement through its capacity to model long-range dependencies and contextual information, yet comparative analysis indicates that QUIET-SR consistently produces superior results across the evaluated datasets while maintaining a relatively efficient model size of $1.55$MB, potentially making it suitable for deployment in environments with computational constraints, with its apparent capacity to preserve structural details being particularly relevant for specialized applications in medical imaging, where reconstruction fidelity can influence diagnostic accuracy and pattern recognition tasks.

\begin{figure}[htpb]
    \centering
    \includegraphics[width=\linewidth]{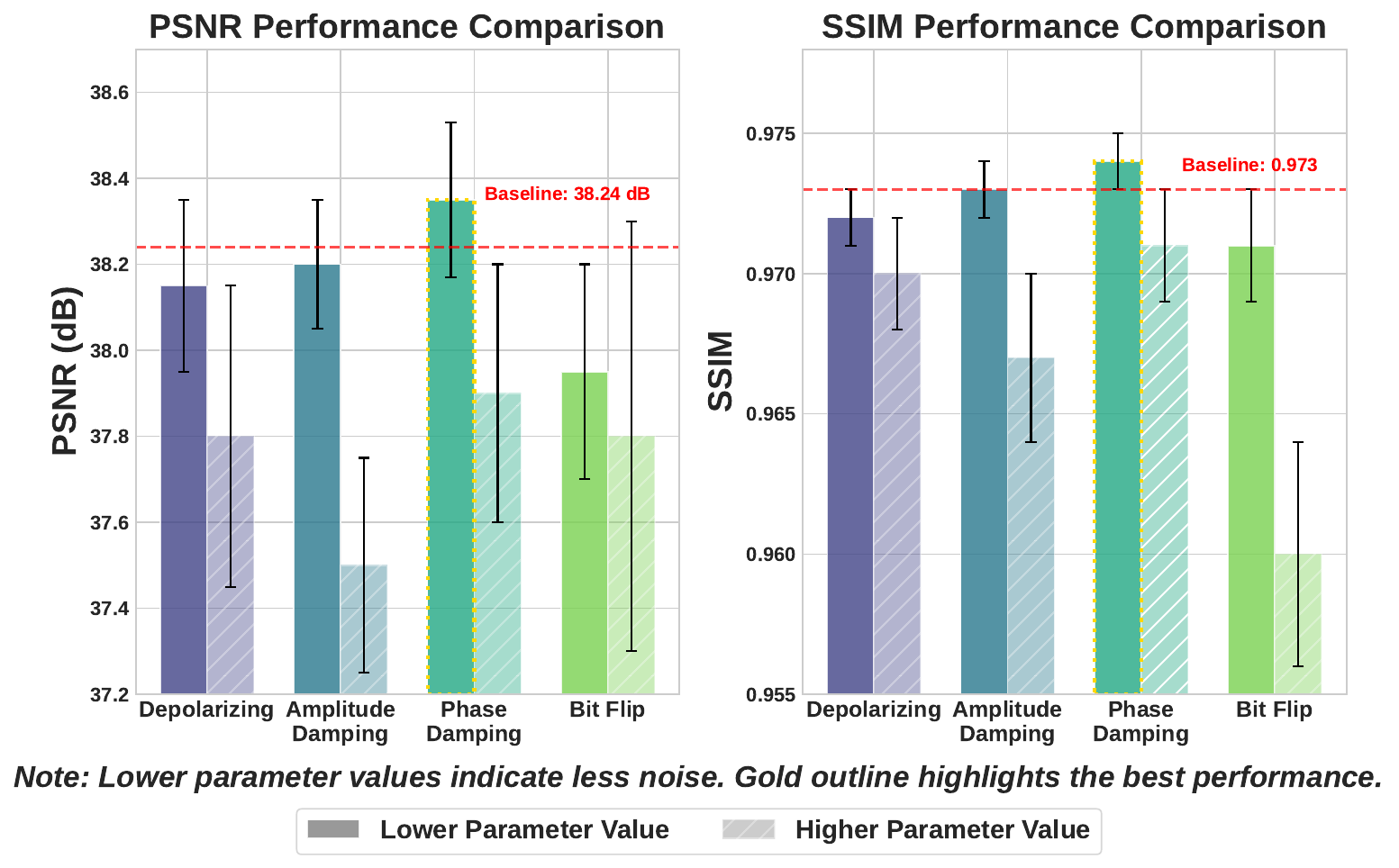}
    \caption{Noise-aware simulation results using Qiskit Aer with Depolarizing, Amplitude Damping, Phase Damping, \& BitFlip noise models. Under low Phase Damping noise, QUIET-SR achieves PSNR $\approx 38.5$ dB \& SSIM $\approx 0.974$, exceeding the noiseless baseline (PSNR $38.24$ dB, SSIM $0.973$). Performance remains close to baseline under Depolarizing noise (PSNR $\approx 38.15$ dB, SSIM $\approx 0.972$), indicating noise resilience.}
    \label{fig:noise}
\end{figure}

\begin{figure}[htbp]
    \centering
    \includegraphics[width=\linewidth]{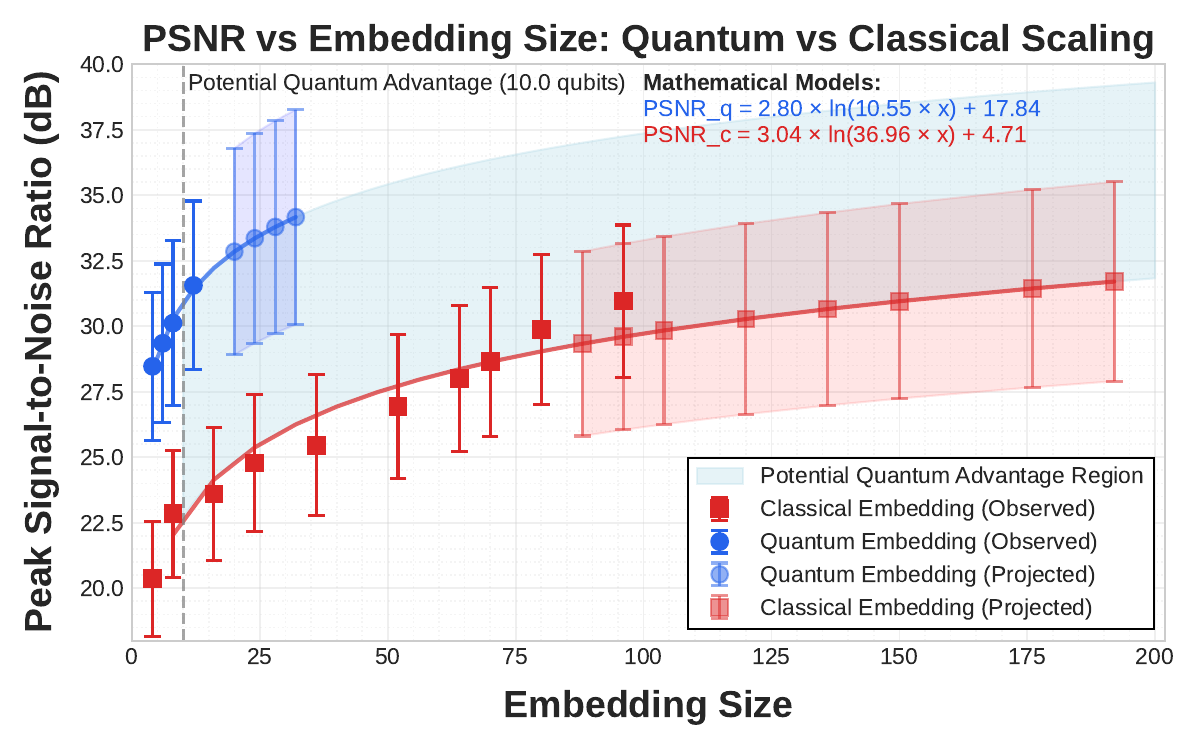}
    \caption{ \small This comparative analysis illustrates the relationship between embedding dimension (number of qubits in quantum systems) and PSNR. Empirical measurements (solid lines) are shown for both quantum and classical embeddings up to 10 qubits, beyond which projections (dashed lines) are generated using a logarithmic regression model. The shaded region represents the predicted performance advantage of QUIET-SR quantum embeddings over their classical counterparts as quantum hardware capabilities expand. The diverging trajectories suggest that quantum embeddings may offer increasingly significant advantages in image reconstruction quality as larger quantum systems become available, with the performance gap widening in proportion to system size.}
    \label{potential}
\end{figure}

\textbf{Why Does QUIET-SR Work Effectively? }\textit{Effect of Quantum Attention Mechanism. }Recalling that QUIET-SR employs a high representational rich embedding feature space provided by the quantum attention mechanism with the goal to effectively capture features that can output a high resolution image from a low resolution input, the comparative analyses in Tables~\ref{tab:medical-comparison1} and \ref{tab:medical-comparison2} offer further insights into its efficacy. The observed performance improvements can be attributed to the quantum attention mechanism’s ability to exploit non-local correlations and capture complex, high-dimensional feature interactions. This is achieved through an adaptive encoding strategy that harnesses multiple rotation bases and entanglement operations, enabling a more discriminative and context-aware representation. Consequently, QUIET-SR is able to preserve fine structural details and achieve superior reconstruction quality across varying imaging scenarios, a result that substantiates its potential in addressing the challenges inherent in super-resolution tasks for medical images. 



\subsection*{Feasibility on NISQ Devices}
We validated the hardware feasibility of QUIET-SR through noise-aware simulations, confirming robustness against realistic quantum noise. These results align with the theoretical propagation of bounded errors \cite{sultanow2024quantum} \& support the practicality of our shallow VQC design for NISQ hardware. Moreover, error mitigation frameworks like \textit{mitiq} \& advances in surface code error correction are expected to further improve performance on quantum devices.

\subsection*{QUIET-SR Resource Efficiency and Embedding Dimension scaling on NISQ-Era Quantum Machines.} As demonstrated in Fig. \ref{potential}, QUIET-SR's performance improvements underscore its scalability on current NISQ quantum machines. Despite qubits remaining a scarce resource in the NISQ era, the distributed architecture of QUIET-SR's quantum circuits enables efficient resource allocation. This design facilitates parallel execution of multiple circuits, allowing scaling to larger embedding dimensions even with limited qubit availability. The performance curve indicates that even modest increases in embedding size yield noticeable gains in PSNR, reflecting improved recovery of high-resolution details. This behavior demonstrates that the variational quantum circuits within QUIET-SR capture complex, non-local features in image data that traditional methods might overlook. As a result, QUIET-SR establishes itself as the first variational image super-resolution algorithm leveraging distributed quantum processing via efficiently utilizing limited qubit resources while establishing a promising direction for future developments as quantum hardware advances.

\section{Related Work \label{sec2}}

\textbf{Limitations of Quantum Computing in SISR and the Opportunity Gap:}  
The integration of QC into SISR is an emerging research area. One early approach recasts the sparse coding optimization problem as a Quadratic Unconstrained Binary Optimization (QUBO) problem \cite{7966350}, expressed as:
\begin{equation}
\min_{\mathbf{z} \in \{0,1\}^n} \left( \mathbf{z}^T \mathbf{Q} \mathbf{z} + \mathbf{c}^T \mathbf{z} \right),
\end{equation}
where \( \mathbf{Q} \) is a matrix that encodes the interaction between candidate basis elements, \( \mathbf{z} \) is a binary vector representing the selection of these elements, and \( \mathbf{c} \) accounts for linear contributions. Preliminary experiments using D-Wave's 5760-qubit quantum annealer have shown promise in solving these QUBO formulations, potentially enabling faster discovery of sparse representations that aid HR reconstruction \cite{choong2023quantum}. However, no implementations have yet leveraged current NISQ quantum computers for image super-resolution, highlighting an opportunity gap and the need for techniques to make it feasible.

\textbf{Deep Learning Approaches in SISR:}
The evolution of deep learning in SISR began with SRCNN, introduced by Dong et \textit{al.} \cite{dong2015image}, which established the foundational CNN framework for super-resolution. This model operated in three key stages: patch extraction, where small overlapping regions of the low-resolution image were sampled as input; non-linear mapping, where a deep network learned complex transformations to infer high-resolution details; and reconstruction, where these enhanced patches were combined to generate the final high-resolution output. A major breakthrough came with SRGAN by Ledig et \textit{al.} \cite{ledig2017photo}, which introduced GANs for SISR. By incorporating perceptual and adversarial loss functions, SRGAN focused on producing images with more realistic textures and finer details, moving beyond traditional pixel-wise optimization. Further advancements, such as EDSR \cite{lim2017enhanced} and RCAN \cite{zhang2018image}, refined super-resolution by introducing enhanced residual learning and channel attention mechanisms, allowing deep networks to focus more effectively on key image features and improving their ability to learn complex high-resolution-to-low-resolution mappings.

\textbf{Sparse Approaches in SISR:}  
Sparse coding methods have provided a strong foundation for super-resolution, with Yang et \textit{al.} \cite{4587647} demonstrating the effectiveness of learning structured mappings between LR and HR image patches. Instead of directly mapping an LR patch to its HR counterpart, sparse coding represents each LR patch as a weighted combination of a small set of fundamental patterns (basis vectors), which are shared between the LR and HR domains. These sets of patterns, known as dictionaries, allow the model to reconstruct HR images using the same sparse representation \( \boldsymbol{\alpha} \) found from the LR image. Mathematically, this is expressed as:  
\begin{equation}
\mathbf{x} \approx \mathbf{D}_{\text{LR}} \boldsymbol{\alpha} \quad \text{and} \quad \mathbf{y} \approx \mathbf{D}_{\text{HR}} \boldsymbol{\alpha},
\end{equation}  
where \( \mathbf{x} \) is the LR patch, \( \mathbf{y} \) is the reconstructed HR patch, \( \mathbf{D}_{\text{LR}} \) and \( \mathbf{D}_{\text{HR}} \) are the learned LR and HR dictionaries (i.e., sets of representative patterns), and \( \boldsymbol{\alpha} \) is a sparse coefficient vector indicating which patterns are combined to approximate the image patch. The optimization problem for finding the best \( \boldsymbol{\alpha} \) is formulated as:  
\begin{equation}
\min_{\boldsymbol{\alpha}} \| \mathbf{x} - \mathbf{D}_{\text{LR}} \boldsymbol{\alpha} \|_2^2 + \lambda \| \boldsymbol{\alpha} \|_1,
\end{equation}  
where \( \lambda \) is a regularization parameter that encourages sparsity, ensuring that only a few basis vectors are selected for reconstruction. Timofte et \textit{al.} \cite{6751349} later improved upon this approach with anchored neighborhood regression, which combines local similarity-based regression with precomputed global transformations. This method significantly speeds up computation by replacing the iterative sparse coding step with fast lookup-based matrix operations.

\textbf{Attention Mechanisms used in SISR:}  
Attention mechanisms have significantly advanced SISR by enabling models to selectively emphasize image features that are most relevant for reconstructing high-quality details at different scales. These mechanisms help the network focus on fine textures, edges, and structural patterns that are often lost in low-resolution images.
One such approach is the Pixel Attention Network (PAN) by Zhao et al. \cite{zhao2020efficient}, which introduces a method to dynamically adjust the importance of individual pixels based on their contribution to the overall image structure. This is achieved through an attention map, which assigns an adaptive weight to each pixel in the feature representation, determining how much influence it should have in the reconstruction process. Mathematically, this is represented by a three-dimensional matrix \( A \in \mathbb{R}^{C \times H \times W} \), where \( C \) is the number of feature channels, and \( H \) and \( W \) are the spatial dimensions. The enhanced feature map \( x_k \) after applying pixel attention is given by:
\begin{equation}
x_k = f_{PA}(x_{k-1}) \cdot x_{k-1},
\end{equation}
where \( f_{PA}(\cdot) \) represents a \( 1 \times 1 \) convolution followed by a sigmoid activation function. This step refines the feature representation by selectively amplifying important pixel contributions and suppressing less informative ones.

Transformer-based architectures, such as the Swin Transformer \cite{liang2021swinir, conde2022swin2sr}, build upon traditional attention mechanisms by organizing image information in a structured manner. Instead of analyzing the entire image at once, these models break it into smaller, non-overlapping regions, referred to as windows, and process them separately. To ensure the network still captures larger structures, the position of these windows is shifted in subsequent layers, allowing different parts of the image to be connected progressively. This hierarchical approach allows the model to first focus on fine details within small regions before gradually integrating broader patterns across the entire image. By balancing local precision with a global understanding of the image, this method significantly improves reconstruction quality. However, these models often require a large number of parameters, which increases computational costs, motivating the exploration of alternative approaches that can achieve similar performance while reducing complexity.
\begin{figure*}[t]
    \centering
    \includegraphics[width=1\linewidth]{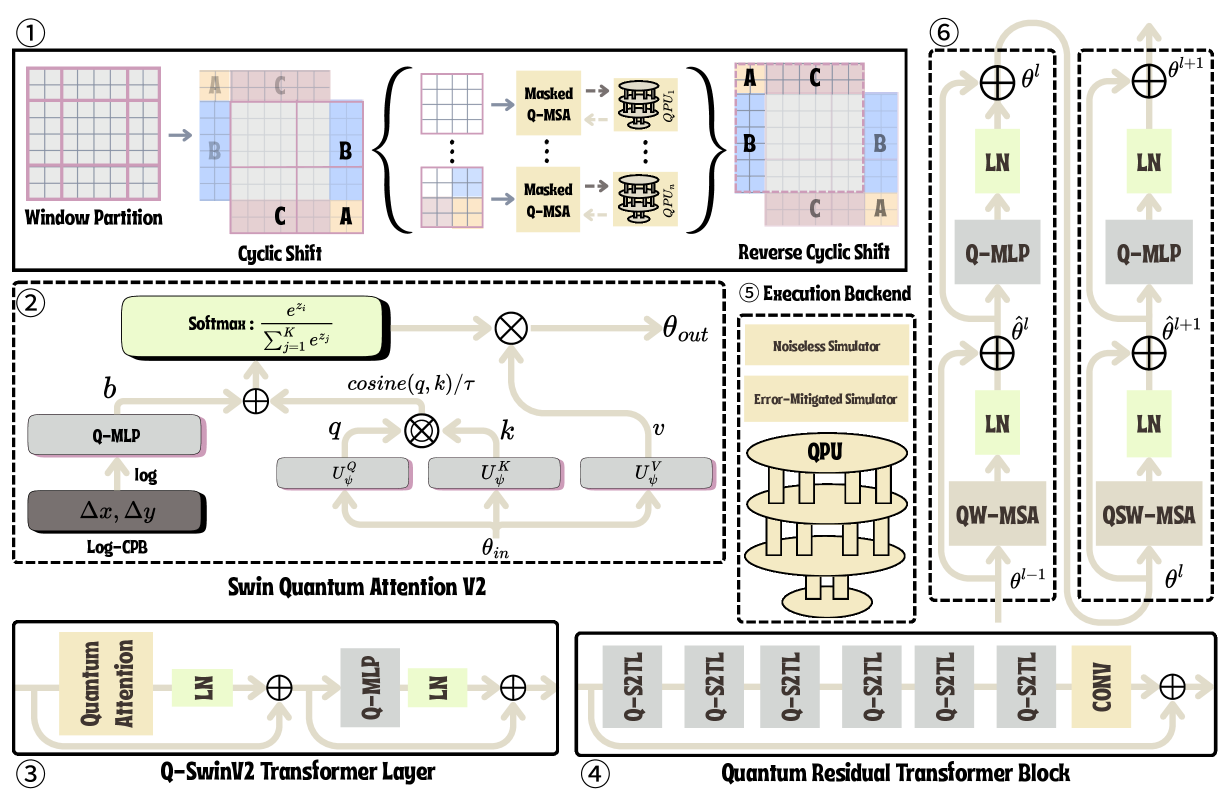}
    \caption{ \small \textbf{Detailed architecture of the core QUIET-SR components.} \textbf{(1)} The \textbf{Shifted Quantum Window} mechanism utilizes cyclic shifting and masking to partition the input into windows, enabling efficient cross-window interaction on quantum processors which can be distributed. \textbf{(2)} The \textbf{Swin Quantum Attention V2} module replaces classical linear projections with variational quantum circuits ($U_\psi^Q, U_\psi^K, U_\psi^V$) and incorporates a \textbf{Log-CPB} (Log-spaced Continuous Position Bias) processed by a Q-MLP to compute scaled cosine attention. \textbf{(3)} The \textbf{Q-SwinV2 Transformer Layer} integrates the quantum attention mechanism with a Quantum MLP (Q-MLP), Layer Normalization (LN), and residual connections. \textbf{(4)} The \textbf{Quantum Residual Transformer Block} stacks multiple transformer layers (Q-S2TL) followed by a convolutional layer. \textbf{(5)} The \textbf{Execution Backend} executes the compiled quantum circuits on the chosen backend: a noiseless simulator, an error-mitigated simulator, or a real quantum processing unit (QPU). \textbf{(6)} The alternating processing strategy between Quantum Window (QW-MSA) and Quantum Shifted Window (QSW-MSA) attention layers, which facilitates global information flow and feature mixing.}
    \label{framework}
\end{figure*}
\section{Proposed Approach and Key Ideas\label{sec3}}
In this section, we present the design and implementation of QUIET-SR, a quantum image enhancement transformer for SISR as shown in Fig. \ref{framework}. Our approach integrates quantum-enhanced attention mechanisms within a classical deep learning framework to efficiently reconstruct high-resolution images from low-resolution inputs. This section outlines the architecture of QUIET-SR, including the key quantum components and their role in image reconstruction.

\textbf{Variational Quantum Architecture.} QUIET-SR employs a variational/hybrid quantum-classical architecture for SISR as visually portrayed in Fig. \ref{fig:comparision}. The quantum components appear in two specific areas: the MLP and the attention mechanism, which are replaced by Quantum MLP and quantum attention mechanism respectively. The remainder of the architecture maintains its classical nature, allowing for the utilization of classical computing's data processing capabilities in image processing. There are key novel architectural implementations that highlight the improvements: (1) to address the scalability and quantum resource bottlenecks for generating high-quality super-resolution images, QUIET-SR utilizes classical patch embeddings by dividing the input image \( I \) into smaller non-overlapping patches \(P\_i\), and (2) to reduce parameterization caused by classical attention layers in the shifted window attention mechanism of Swin Transformer, we utilize a quantum attention mechanism\cite{dutta2024aqpinns, dutta2024qadqn} which leverages the probabilistic states of qubits as queries, keys, and values within the Hilbert space to effectively capture the interrelated relationships within the pixel-wise information.

\textbf{Patch Embedding and Shallow Feature Extraction.} A key design decision in our framework stems from the limited number of qubits available, making it challenging to process images in their entirety.  We address the limited number of qubits available by dividing the input image \( I \) into smaller non-overlapping patches \( P_i \), where each patch is represented as:
\begin{equation}
P_i = I \left[ h:h+p, w:w+p \right], \quad i \in \{1, 2, ..., N \},
\end{equation}
where \( p \times p \) is the patch size, \( (h, w) \) represents the starting position of the patch, and \( N \) is the total number of patches. To extract shallow features, we apply a convolutional layer. Let \( F_{\text{shallow}} \) denote the feature map obtained after convolution. These extracted shallow features \( F_{\text{shallow}} \) retain low-frequency information \cite{conde2022swin2sr} and serve as input for subsequent quantum attention mechanisms.

\textbf{Quantum Multi-layer Perceptron.} This architecture integrates quantum processing capabilities within a classical neural network framework to enhance feature representation and transformation. The QMLP extends traditional MLPs by incorporating a variational quantum circuit as a core computational element between classical linear transformations. The transformation process in QMLP can be formally expressed through a sequence of operations. The quantum processing layer employs a multi-basis rotation encoding scheme that varies depending on the depth of the quantum circuit. For deeper quantum circuits where the number of layers are more than $1$, the system utilizes all three Pauli rotation gates ($R_X$, $R_Y$, and $R_Z$), while for shallow circuits where there is monolayer circuit, only $R_Z$ rotations are applied. This adaptive approach can be formalized as:
\begin{equation}
\mathcal{R} = \begin{cases}
\{R_X, R_Y, R_Z\}, & \text{if $L$} > 1, \\
\{R_Z\}, & \text{if $L$} = 1,
\end{cases}
\end{equation}
where $N_{L}$ are the number of quantum layers. 

For each rotation basis $R \in \mathcal{R}$, the input features $\mathbf{h}_1$ are encoded into the quantum system through angle embedding:
\begin{equation}
S_R(\mathbf{h}_1) = \prod_{i=1}^{n} R(\phi_i(\mathbf{h}_1)),
\end{equation}
where $\phi_i(\mathbf{h}_1)$ represents the mapping of classical data to rotation angles for the $i$-th qubit, and $R \in \{R_X, R_Y, R_Z\}$ corresponds to the rotation operators around the $X$, $Y$, and $Z$ axes respectively:

\[
\scalebox{0.8}{$
R(\vec{n}, \phi) =
\begin{pmatrix}
\cos(\phi/2) - i n_z \sin(\phi/2) & -i (n_x - i n_y) \sin(\phi/2) \\
-i (n_x + i n_y) \sin(\phi/2) & \cos(\phi/2) + i n_z \sin(\phi/2)
\end{pmatrix}
$}
\]

where \( \vec{n} = (n_x, n_y, n_z) \) is a unit vector defining the rotation axis, and \( \vec{\sigma} = (X, Y, Z) \) represents the Pauli matrices. Following each rotation encoding, an entangling layer is applied:
\begin{equation}
V(\boldsymbol{\theta}) = \prod_{l=1}^{L} \left[ \prod_{i=1}^{n} R_{\phi}(\theta_{l,i}) \right] \left[ \prod_{i=1}^{n-1} \text{CNOT}_{i,i+1} \right],
\end{equation}
where $\text{CNOT}_{i,j}$ represents a controlled-NOT gate with qubit $i$ as control and qubit $j$ as target, enabling quantum entanglement between adjacent qubits. This entanglement operation is crucial for establishing non-local correlations that contribute to the quantum advantage in feature processing. The complete quantum state preparation combines these rotational encodings with entangling layers:
\begin{equation}
|\psi(\mathbf{h}_1, \boldsymbol{\theta})\rangle = \prod_{R \in \mathcal{R}} V(\boldsymbol{\theta}_R) S_R(\mathbf{h}_1) |\mathbf{0}\rangle^{\otimes n}.
\end{equation}
The measurement process extracts classical information from the quantum state through expectation values of appropriate observables. The choice of observables depends on the quantum circuit depth:
\begin{equation}
\mathcal{O} = \begin{cases}
\{X_i, Y_i, Z_i\}_{i=1}^{n}, & \text{if $L$} > 1, \\
\{Z_i\}_{i=1}^{n}, & \text{if $L$} = 1,
\end{cases}
\end{equation}
where $X_i$, $Y_i$, and $Z_i$ represent the Pauli operators applied to the $i$-th qubit. The expectation values are computed as:
\begin{equation}
[\mathbf{h}_2]_j = \langle\psi(\mathbf{h}_1, \boldsymbol{\theta})|O_j|\psi(\mathbf{h}_1, \boldsymbol{\theta})\rangle,
\end{equation}
for each observable $O_j \in \mathcal{O}$. This multi-basis rotation scheme provides several theoretical advantages. For deeper circuits, using all three rotation bases ($X$, $Y$, $Z$) allows the quantum system to explore a more complete Hilbert space, enabling more complex transformations. For shallow circuits, focusing solely on $Z$ rotations provides computational efficiency while still leveraging quantum effects for feature processing.

\textbf{Efficient Parallel Quantum Processing via Cyclic Shifting in a Quantum Supercomputing Paradigm.} A bottleneck in hybrid quantum-classical networks is the communication overhead and latency associated with accessing Quantum Processing Units (QPUs) when individual circuit instances scale up quickly due to complexity. To address this, inspired by the efficient batch computation strategy for graphic processing units (GPUs) introduced in the Swin Transformer paper~\cite{liang2021swinir}, QUIET-SR implements an efficient batch computation procedure facilitated by a cyclic shifting scheme, as illustrated in Part 1 of Fig.~\ref{framework}. In standard window-based attention, processing edge tiles often requires padding, which introduces non-uniform data structures that complicate quantum circuit compilation.

By cyclically shifting the image feature map toward the top-left direction, we ensure that the image is partitioned into uniform, fixed-size windows without the need for padding. This uniformity is critical for quantum hardware integration; it allows disparate window processing tasks to be structured as a single batch of identical quantum circuits with the same ansatz and qubit requirements. To handle the semantic discontinuities introduced by the shift, where non-adjacent sub-windows are brought together. We employ a masked attention mechanism that inhibits information flow between unconnected regions during the quantum state evolution.

This batching strategy directly enables massive parallelization on NISQ hardware. Since the quantum circuits for each window are topologically identical and independent, they can be distributed across multiple distributed QPUs or executed in parallel on different regions of a larger quantum processor without the need for circuit reconfiguration. This parallel execution paradigm significantly mitigates the latency bottlenecks inherent in sequential quantum state preparation and measurement, ensuring that the framework remains computationally viable for high-resolution inputs and each individual circuit instance's size scales slower with complexity. This protocol aligns with emerging quantum supercomputing paradigms based on coordinated multi-QPU–GPU infrastructures \cite{mohseni2025}.

\begin{algorithm}[H]
\caption{QUIET-SR}
\label{alg:qsr}
\footnotesize
\KwIn{Low-resolution image $I_{LR} \in \mathbb{R}^{H \times W \times C}$, Upscaling factor $s$, Window size $M$, Number of quantum layers $L$} 
\KwOut{High-resolution image $I_{HR} \in \mathbb{R}^{sH \times sW \times C}$}

\textbf{Shallow Feature Extraction:} \\
$F_0 \gets \text{Conv}_{3\times3}(I_{LR})$ 

\SetKwFunction{FQuantumLayer}{QuantumLayer}
\SetKwFunction{FShiftedQuantumWindowAttention}{ShiftedQuantumWindowAttention}

\SetKwProg{Fn}{Function}{:}{}
\Fn{\FQuantumLayer{$x, \theta$}}{
    Initialize Pennylane quantum circuit with $n$ qubits \\
    Encode input $x$ into quantum state using rotational embeddings \\
    Apply parameterized quantum layers with trainable weights $\theta$ \\
    Measure expectation values: $\langle Z_i \rangle$ for $i \in [1,n]$ \\
    \KwRet $[\langle Z_1 \rangle, \dots, \langle Z_n \rangle]$
}

\Fn{\FShiftedQuantumWindowAttention{$X \in \mathbb{R}^{B \times N \times D}, \text{shift\_size}$}}{
    Partition input into windows: $\{W_i\}_{i=1}^{N_w}$ of size $M \times M$ 
    
    \If{$\text{shift\_size} > 0$}{
        Apply circular shift: $X_{\text{shifted}} \gets \text{Roll}(X, \text{shifts}=(-\text{shift\_size}, -\text{shift\_size}))$
    }
    \Else{
        $X_{\text{shifted}} \gets X$
    }
    
    \ForEach{$W_i$}{
        $QKV_i \gets \text{QuantumLayer}(W_i*3, \theta_{QKV})$ 
        
        Compute quantum-enhanced attention:
        $A_i \gets \frac{Q_i K_i^T}{\sqrt{d}} + B$, where $B$ is the relative position bias
        
        Apply softmax: $A_i \gets \text{softmax}(A_i)$
        
        Compute output projection: 
        $O_i \gets A_i V_i$ 
        
        Apply final QuantumLayer: 
        $O_i \gets \text{QuantumLayer}(O_i, \theta_O)$
    }
    
    \If{$\text{shift\_size} > 0$}{
        Reverse the shift: 
        $O_{\text{final}} \gets \text{Roll}(O, \text{shifts}=(\text{shift\_size}, \text{shift\_size}))$
    }
    
    \KwRet{$O_{\text{final}}$}
}

\textbf{Main Network Forward Pass:} \\
\For{$l \gets 1$ \textbf{to} $L$}{
    \textbf{Quantum Swin Transformer Block:} \\
    $X_l \gets \text{LayerNorm}(F_{l-1})$ \\
    $A_l \gets \text{ShiftedQuantumWindowAttention}(X_l, \text{shift\_size}=l \bmod 2 \times M / 2)$ \\
    $F'_l \gets F_{l-1} + \text{DropPath}(A_l)$

    \textbf{Quantum MLP Block:} \\
    $Y_l \gets \text{LayerNorm}(F'_l)$ \\
    $M_l \gets \text{QuantumLayer}(\text{MLP}(Y_l), \theta_{MLP})$ \\
    $F_l \gets F'_l + \text{DropPath}(M_l)$
}

\textbf{Upsampling:} \\
$F_{\text{up}} \gets \text{PixelShuffle}(\text{Conv}_{3\times3}(F_L))$ \\
$I_{HR} \gets \text{Conv}_{3\times3}(F_{\text{up}})$

\KwRet{$I_{HR}$}
\end{algorithm}

\textbf{Shifted Quantum Window Attention (SQWIN).} This mechanism combines principles of self-attention and quantum variational circuits \cite{benedetti2019parameterized} to process image features efficiently, as illustrated in the architecture shown in the subsection $D$ of Fig.~\ref{framework}. The quantum state is initially prepared using the parameterized quantum circuit, where the state vector is defined as:
\begin{equation}
    \ket{\psi(\theta)} = M(\theta) \ket{0}^{\otimes n},
\end{equation}
where \( M(\theta) \) represents the parameterized quantum circuit, and \( n \) denotes the number of qubits. The quantum circuit consists of sequential layers, where each layer involves rotation gates \( R_X(\theta_{l,i}) \) applied to each qubit, followed by entangling operations using controlled-NOT (CNOT) gates between consecutive qubits. The full circuit is described by \(V(\theta)\),
where \( L \) is the number of layers, and the combination of \( R_X(\theta_{l,i}) = e^{-i \theta_{l,i} X} \) and CNOT gates enables the generation of entangled quantum states that are crucial for capturing complex features in images. Our implementation utilizes a QuantumLayer consisting of rotational embeddings followed by entangling operations. The quantum attention mechanism employs cosine similarity with a learnable temperature parameter to compute attention weights, which is then applied in the quantum domain to model interactions between image patches.
\begin{equation}
    \text{Attn}(qkv) = \text{softmax} \left( \kappa \cdot \cos(qkv) + B_{\text{rel}} \right),
\end{equation}
where $\kappa$  is a learnable parameter that controls the attention logits and $B_{\text{rel}}$ represents the relative position bias. This bias is computed through a continuous MLP-based approach $B_{\text{rel}} = \text{QMLP}(R_{\text{table}})$ where $R_{\text{table}}$ contains the normalized relative position coordinates transformed using a logarithmic encoding.  These coordinates are scaled and processed through a two-layer MLP to generate head-specific positional biases. This quantum formulation of cosine attention ensures that the relationships between image patches are effectively captured, enabling the recovery of high-resolution details in the image. The final output is computed, where a second quantum transformation is applied to the weighted combination of values. This dual quantum processing enables the model to better capture non-local dependencies and complex feature relationships, enhancing the recovery of high-resolution details in images.

\textbf{Quantum Advantage in Attention Computation:} Classical scaled dot-product attention requires computing inner products \( s_j = \mathbf{q} \cdot \mathbf{k}_j \) for \(M\) keys of dimension \(N\), with complexity \(O(MN)\). On a QC, these inner products can be estimated via the swap test in time $O\left(M \cdot\frac{\mathrm{polylog}(N)}{\varepsilon}\right)$
where
$
\ket{\psi_{\mathbf{q}}} = \frac{1}{\|\mathbf{q}\|} \sum_{i=1}^N q_i \ket{i}, \quad \ket{\psi_{\mathbf{k}_j}} = \frac{1}{\|\mathbf{k}_j\|} \sum_{i=1}^N k_{j,i} \ket{i},
$
\& \(\varepsilon\) is the additive estimation error. Classical methods must read all \(N\) components per key, imposing a lower bound of \(\Omega(MN)\). Hence, quantum attention offers a polynomial speedup in \(N\), enabling faster similarity estimation.

\textbf{Quantum Resource Estimation of QUIET-SR.} For each quantum layer in the QISR model, the resource requirements are determined by the number of qubits needed to encode the quantum states and any additional ancilla qubits required for the quantum circuit implementation $\lceil \log_2(D) \rceil + a$, where \( \lceil \log_2(D) \rceil \) represents the number of qubits required to encode the \( D \)-dimensional quantum state, where \( D \) is the embedding dimension. The term \( a \) denotes the additional ancilla qubits necessary to facilitate the quantum computation, such as those used for error correction or intermediate calculations. The quantum circuit depth scales as \( \mathcal{O}(D\log D) \) due to the quantum attention mechanism, which involves operations that grow logarithmically with respect to \( D \). This scaling arises from the quantum attention mechanism, where the quantum states interact and produce results that depend on these quantum dimensions. The quantum state preparation itself is represented as a linear combination of quantum basis states $|\psi(x)\rangle = \sum_{i=0}^{D-1} \alpha_i|i\rangle$  where \( \alpha_i \) are the normalized feature values extracted from the input features. The relative position bias term \( B \) is computed using a combination of learned positional encoding terms:
\[
\scalebox{0.9}{$
B_{i,j} = \text{QMLP}\!\left(
\log_2\!\left(1 + \frac{|\Delta x|}{\gamma_x}\right)
\oplus
\log_2\!\left(1 + \frac{|\Delta y|}{\gamma_y}\right)
\right)
$}
\]
where \( \Delta x \) and \( \Delta y \) represent the differences in positions between elements, and \( \gamma_x \) and \( \gamma_y \) are learnable parameters. These terms contribute to capturing positional relationships between features, improving the attention mechanism.
\section{Conclusion\label{sec5}}
QUIET-SR is the first hybrid quantum-classical framework that demonstrates the practical potential of QC in image processing applications within current hardware constraints. By successfully implementing a quantum-enhanced super-resolution system that operates within a limited qubit environment constraint per circuit, we have shown that meaningful quantum advantages can be achieved even with NISQ devices.

\begin{acknowledgments}
This work was supported in part by the NYUAD Center for Quantum and Topological Systems (CQTS), funded by
Tamkeen under the NYUAD Research Institute grant CG008.
\end{acknowledgments}


\bibliographystyle{apsrev4-1-title}
\bibliography{main}

@String(ECCV= {Eur. Conf. Comput. Vis.})

@String(ECCV  = {ECCV})

@article{chen2024single,
  title={Single image super-resolution based on trainable feature matching attention network},
  author={Chen, Qizhou and Shao, Qing},
  journal={Pattern Recognition},
  volume={149},
  pages={110289},
  year={2024},
  publisher={Elsevier}

}

@Article{rs14215423,
AUTHOR = {Wang, Xuan and Yi, Jinglei and Guo, Jian and Song, Yongchao and Lyu, Jun and Xu, Jindong and Yan, Weiqing and Zhao, Jindong and Cai, Qing and Min, Haigen},
TITLE = {A Review of Image Super-Resolution Approaches Based on Deep Learning and Applications in Remote Sensing},
JOURNAL = {Remote Sensing},
VOLUME = {14},
YEAR = {2022},
NUMBER = {21},
ARTICLE-NUMBER = {5423},
URL = {https://www.mdpi.com/2072-4292/14/21/5423},
ISSN = {2072-4292},
DOI = {10.3390/rs14215423}
}

@INPROCEEDINGS{7095900,
  author={Isaac, Jithin Saji and Kulkarni, Ramesh},
  booktitle={2015 International Conference on Technologies for Sustainable Development (ICTSD)}, 
  title={Super resolution techniques for medical image processing}, 
  year={2015},
  volume={},
  number={},
  pages={1-6},
  doi={10.1109/ICTSD.2015.7095900}}

@InProceedings{10.1007/978-3-319-10662-5_18,
author="Okarma, Krzysztof
and Tec{\l}aw, Mateusz
and Lech, Piotr",
editor="Chora{\'{s}}, Ryszard S.",
title="Application of Super-Resolution Algorithms for the Navigation of Autonomous Mobile Robots",
booktitle="Image Processing {\&} Communications Challenges 6",
year="2015",
publisher="Springer International Publishing",
address="Cham",
pages="145--152",
isbn="978-3-319-10662-5"
}

@article{dong2015image,
  title={Image super-resolution using deep convolutional networks},
  author={Dong, Chao and Loy, Chen Change and He, Kaiming and Tang, Xiaoou},
  journal={IEEE transactions on pattern analysis and machine intelligence},
  volume={38},
  number={2},
  pages={295--307},
  year={2015},
  publisher={IEEE}
}

@inproceedings{lim2017enhanced,
  title={Enhanced deep residual networks for single image super-resolution},
  author={Lim, Bee and Son, Sanghyun and Kim, Heewon and Nah, Seungjun and Mu Lee, Kyoung},
  booktitle={Proceedings of the IEEE conference on computer vision and pattern recognition workshops},
  pages={136--144},
  year={2017}
}

@inproceedings{zhang2018image,
  title={Image super-resolution using very deep residual channel attention networks},
  author={Zhang, Yulun and Li, Kunpeng and Li, Kai and Wang, Lichen and Zhong, Bineng and Fu, Yun},
  booktitle={Proceedings of the European conference on computer vision (ECCV)},
  pages={286--301},
  year={2018}
}

@inproceedings{ledig2017photo,
  title={Photo-realistic single image super-resolution using a generative adversarial network},
  author={Ledig, Christian and Theis, Lucas and Husz{\'a}r, Ferenc and Caballero, Jose and Cunningham, Andrew and Acosta, Alejandro and Aitken, Andrew and Tejani, Alykhan and Totz, Johannes and Wang, Zehan and others},
  booktitle={Proceedings of the IEEE conference on computer vision and pattern recognition},
  pages={4681--4690},
  year={2017}
}

@inproceedings{liang2021swinir,
  title={Swinir: Image restoration using swin transformer},
  author={Liang, Jingyun and Cao, Jiezhang and Sun, Guolei and Zhang, Kai and Van Gool, Luc and Timofte, Radu},
  booktitle={Proceedings of the IEEE/CVF international conference on computer vision},
  pages={1833--1844},
  year={2021}
}

@article{wang2022review,
  title={Review of quantum image processing},
  author={Wang, Zhaobin and Xu, Minzhe and Zhang, Yaonan},
  journal={Archives of Computational Methods in Engineering},
  volume={29},
  number={2},
  pages={737--761},
  year={2022},
  publisher={Springer}
}

@inproceedings{choong2023quantum,
  title={Quantum annealing for single image super-resolution},
  author={Choong, Han Yao and Kumar, Suryansh and Van Gool, Luc},
  booktitle={Proceedings of the IEEE/CVF Conference on Computer Vision and Pattern Recognition},
  pages={1150--1159},
  year={2023}
}

@article{biamonte2017quantum,
  title={Quantum machine learning},
  author={Biamonte, Jacob and Wittek, Peter and Pancotti, Nicola and Rebentrost, Patrick and Wiebe, Nathan and Lloyd, Seth},
  journal={Nature},
  volume={549},
  number={7671},
  pages={195--202},
  year={2017},
  publisher={Nature Publishing Group UK London}
}

@article{preskill2018quantum,
  title={Quantum computing in the NISQ era and beyond},
  author={Preskill, John},
  journal={Quantum},
  volume={2},
  pages={79},
  year={2018},
  publisher={Verein zur F{\"o}rderung des Open Access Publizierens in den Quantenwissenschaften}
}

@inproceedings{conde2022swin2sr,
  title={Swin2sr: Swinv2 transformer for compressed image super-resolution and restoration},
  author={Conde, Marcos V and others },
author2={and Choi, Ui-Jin and Burchi, Maxime and Timofte, Radu},
  booktitle={European Conference on Computer Vision},
  pages2={669--687},
  year={2022},
  organization={Springer}
}

@INPROCEEDINGS{7966350,
  author={Alom, Md Zahangir and Van Essen, Brian and Moody, Adam T. and Widemann, David Peter and Taha, Tarek M.},
  booktitle={2017 International Joint Conference on Neural Networks (IJCNN)}, 
  title={Quadratic Unconstrained Binary Optimization (QUBO) on neuromorphic computing system}, 
  year={2017},
  volume={},
  number={},
  pages={3922-3929},
  doi={10.1109/IJCNN.2017.7966350}}

@inproceedings{zhao2020efficient,
  title={Efficient image super-resolution using pixel attention},
  author={Zhao, Hengyuan and Kong, Xiangtao and He, Jingwen and Qiao, Yu and Dong, Chao},
  booktitle={Computer Vision--ECCV 2020 Workshops: Glasgow, UK, August 23--28, 2020, Proceedings, Part III 16},
  pages={56--72},
  year={2020},
  organization={Springer}
}

@INPROCEEDINGS{6751349,
  author={Timofte, Radu and De, Vincent and Gool, Luc Van},
  booktitle={2013 IEEE International Conference on Computer Vision}, 
  title={Anchored Neighborhood Regression for Fast Example-Based Super-Resolution}, 
  year={2013},
  volume={},
  number={},
  pages={1920-1927},
  doi={10.1109/ICCV.2013.241}}

@INPROCEEDINGS{4587647,
  author={Jianchao Yang and Wright, John and Huang, Thomas and Yi Ma},
  booktitle={2008 IEEE Conference on Computer Vision and Pattern Recognition}, 
  title={Image super-resolution as sparse representation of raw image patches}, 
  year={2008},
  volume={},
  number={},
  pages={1-8},
  doi={10.1109/CVPR.2008.4587647}}

@article{benedetti2019parameterized,
  title={Parameterized quantum circuits as machine learning models},
  author={Benedetti, Marcello and Lloyd, Erika and Sack, Stefan and Fiorentini, Mattia},
  journal={Quantum Science and Technology},
  volume={4},
  number={4},
  pages={043001},
  year={2019},
  publisher={IOP Publishing}
}

@article{dutta2024qadqn,
  title={{QADQN}: Quantum attention deep q-network for financial market prediction},
  author={Dutta, Siddhant and Innan, Nouhaila and Marchisio, Alberto and Yahia, Sadok Ben and Shafique, Muhammad},
  journal={arXiv preprint arXiv:2408.03088},
  year={2024}
}

@ARTICLE{1284395,
  author={Zhou Wang and Bovik, A.C. and Sheikh, H.R. and Simoncelli, E.P.},
  journal={IEEE Transactions on Image Processing}, 
  title={Image quality assessment: from error visibility to structural similarity}, 
  year={2004},
  volume={13},
  number={4},
  pages={600-612},
  doi={10.1109/TIP.2003.819861}}

@INPROCEEDINGS{6263880,
  author={Korhonen, Jari and You, Junyong},
  booktitle={2012 Fourth International Workshop on Quality of Multimedia Experience}, 
  title={Peak signal-to-noise ratio revisited: Is simple beautiful?}, 
  year={2012},
  volume={},
  number={},
  pages={37-38},
  doi={10.1109/QoMEX.2012.6263880}}

@article{kingma2014adam,
  title={Adam: A method for stochastic optimization},
  author={Kingma, Diederik P},
  journal={arXiv preprint arXiv:1412.6980},
  year={2014}
}

@article{he2022revisiting,
  title={Revisiting L1 loss in super-resolution: a probabilistic view and beyond},
  author={He, Xiangyu and Cheng, Jian},
  journal={arXiv preprint arXiv:2201.10084},
  year={2022}
}

@article{senokosov2024quantum,
  title={Quantum machine learning for image classification},
  author={Senokosov, Arsenii and Sedykh, Alexandr and Sagingalieva, Asel and Kyriacou, Basil and Melnikov, Alexey},
  journal={Machine Learning: Science and Technology},
  volume={5},
  number={1},
  pages={015040},
  year={2024},
  publisher={IOP Publishing}
}

@article{unlu2024hybrid,
  title={Hybrid quantum vision transformers for event classification in high energy physics},
  author={Unlu, Eyup B and Comajoan Cara, Mar{\c{c}}al and Dahale, Gopal Ramesh and Dong, Zhongtian and Forestano, Roy T and Gleyzer, Sergei and Justice, Daniel and Kong, Kyoungchul and Magorsch, Tom and Matchev, Konstantin T and others},
  journal={Axioms},
  volume={13},
  number={3},
  pages={187},
  year={2024},
  publisher={MDPI}
}

@article{xiao2017fashion,
  title={Fashion-mnist: a novel image dataset for benchmarking machine learning algorithms},
  author={Xiao, Han and Rasul, Kashif and Vollgraf, Roland},
  journal={arXiv preprint arXiv:1708.07747},
  year={2017}
}

@article{deng2012mnist,
  title={The mnist database of handwritten digit images for machine learning research},
  author={Deng, Li},
  journal={IEEE Signal Processing Magazine},
  volume={29},
  number={6},
  pages={141--142},
  year={2012},
  publisher={IEEE}
}

@article{medmnistv2,
    title={MedMNIST v2-A large-scale lightweight benchmark for 2D and 3D biomedical image classification},
    author={Yang, Jiancheng and Shi, Rui and Wei, Donglai and Liu, Zequan and Zhao, Lin and Ke, Bilian and Pfister, Hanspeter and Ni, Bingbing},
    journal={Scientific Data},
    volume={10},
    number={1},
    pages={41},
    year={2023},
    publisher={Nature Publishing Group UK London}
}

@article{comajoan2024quantum,
  title={Quantum Vision Transformers for Quark--Gluon Classification},
  author={Comajoan Cara, Mar{\c{c}}al and Dahale, Gopal Ramesh and Dong, Zhongtian and Forestano, Roy T and Gleyzer, Sergei and Justice, Daniel and Kong, Kyoungchul and Magorsch, Tom and Matchev, Konstantin T and Matcheva, Katia and others},
  journal={Axioms},
  volume={13},
  number={5},
  pages={323},
  year={2024},
  publisher={MDPI}
}

@article{sultanow2024quantum,
  title={Quantum error propagation},
  author={Sultanow, Eldar and others},
authors={Selimllari, Fation and Dutta, Siddhant and Reese, Barry D and Tehrani, Madjid and Buchanan, William J},
  journal={arXiv preprint arXiv:2410.05145},
  year={2024}
}

@inproceedings{gretton2005measuring,
  title={Measuring statistical dependence with Hilbert-Schmidt norms},
  author={Gretton, Arthur and Bousquet, Olivier and Smola, Alex and Sch{\"o}lkopf, Bernhard},
  booktitle={International conference on algorithmic learning theory},
  pages={63--77},
  year={2005},
  organization={Springer}
}

@article{szekely2007measuring,
  title={Measuring and testing dependence by correlation of distances},
  author={Sz{\'e}kely, G{\'a}bor J and Rizzo, Maria L and Bakirov, Nail K},
  year={2007}
}

@misc{mohseni2025,
      title={How to Build a Quantum Supercomputer: Scaling from Hundreds to Millions of Qubits}, 
      author={Masoud Mohseni and Artur Scherer and K. Grace Johnson and Oded Wertheim and Matthew Otten and Navid Anjum Aadit and Yuri Alexeev and Kirk M. Bresniker and Kerem Y. Camsari and Barbara Chapman and Soumitra Chatterjee and Gebremedhin A. Dagnew and Aniello Esposito and Farah Fahim and Marco Fiorentino and Archit Gajjar and Abdullah Khalid and Xiangzhou Kong and Bohdan Kulchytskyy and Elica Kyoseva and Ruoyu Li and P. Aaron Lott and Igor L. Markov and Robert F. McDermott and Giacomo Pedretti and Pooja Rao and Eleanor Rieffel and Allyson Silva and John Sorebo and Panagiotis Spentzouris and Ziv Steiner and Boyan Torosov and Davide Venturelli and Robert J. Visser and Zak Webb and Xin Zhan and Yonatan Cohen and Pooya Ronagh and Alan Ho and Raymond G. Beausoleil and John M. Martinis},
      year={2025},
      eprint={2411.10406},
      archivePrefix={arXiv},
      primaryClass={quant-ph},
      url={https://arxiv.org/abs/2411.10406}, 
}

@inproceedings{dutta2024aqpinns,
  title={AQ-PINNs: Attention-Enhanced Quantum Physics-Informed Neural Networks for Carbon-Efficient Climate Modeling},
  author={Dutta, Siddhant and Innan, Nouhaila and Ben Yahia, Sadok and Shafique, Muhammad},
  booktitle={NeurIPS 2024 Workshop on Tackling Climate Change with Machine Learning},
  url={https://www.climatechange.ai/papers/neurips2024/27},
  year={2024}
}

@article{innan2024financial,
  title={Financial fraud detection using quantum graph neural networks},
  author={Innan, Nouhaila and Sawaika, Abhishek and Dhor, Ashim and Dutta, Siddhant and Thota, Sairupa and Gokal, Husayn and Patel, Nandan and Khan, Muhammad Al-Zafar and Theodonis, Ioannis and Bennai, Mohamed},
  journal={Quantum Machine Intelligence},
  volume={6},
  number={1},
  pages={7},
  year={2024},
  publisher={Springer}
}
\pagebreak

\end{document}